\documentstyle[prl,aps,preprint,12pt]{revtex}

\begin{document}
\author{A. P\'{e}rez-Madrid}
\title{GIBBS ENTROPY AND IRREVERSIBILITY}
\address{Departament de F\'{\i}sica Fonamental,\\
Facultat de F\'{\i}sica, Universitat de Barcelona.\\
Diagonal 647, 08028 Barcelona, Spain}
\maketitle

\begin{abstract}
This contribution is dedicated to dilucidating the role of the Gibbs entropy
in the discussion of the emergence of irreversibility in the macroscopic
world from the microscopic level. By using an extension of the Onsager
theory to the phase space we obtain a generalization of the Liouville
equation describing the evolution of the distribution vector in the form of
a master equation. This formalism leads in a natural way to the breaking of
the BBGKY hierarchy. As a particular case we derive the Boltzmann equation.
\end{abstract}

\pacs{#}


\bigskip

{\em Introduction}.---According to the mechanicistic interpretation of the
physical world, the basic laws of nature are deterministic and time
reversible. However, at the macroscopic level we observe irreversible
processes related to energy degradation, which generate entropy. How do we
reconcile the `spontaneous production of entropy' with the time
reversibility of the microscopic equations of motion?. At the end of
nineteenth century Boltzmann tried to answer this question from a
probabilistic point of view. According to him, entropy is a measure of the
lack of knowledge of the precise state of matter, and can be defined as a
function of the probability of a given state of matter, {\it i.e.} it is a
function of the microstate. All systems in their irreversible evolution tend
to a state of maximum probability or maximum entropy -the state of
equilibrium-.

In contrast to the Boltzmann entropy, the Gibbs entropy is not a function of
the individual microstate but a function of the probability distribution in
a statistical ensemble of systems, both coinciding at equilibrium. As a
consequence of the incompressible character of the flow of points
representing the natural evolution of the statistical ensemble in phase
space , the Gibbs entropy is a constant of motion. Thus, it has been argued
that the relevant entropy for understanding thermodynamic irreversibility is
the Boltzmann entropy and not the Gibbs entropy \cite{lebowitz2}, \cite
{lebowitz}, \cite{goldstein}.

But in our opinion, this is not the end of the story. It is our contention
here to show that depending on the representation we use to describe the
state of the system, the Gibbs entropy is a good definition of the
nonequilibrium entropy which increases in the approach to equilibrium and is
compatible with the Boltzmann's account for irreversibility . To accomplish
that goal, we will use the methods of mesoscopic nonequilibrium
thermodynamics MNET\cite{agusti}, \cite{mazur} in addition to the
generalized Liouville description of dynamical systems in terms of the
distribution vector.

The paper is organized as follows. In Section {\bf II} we introduce the
representation of the state of the isolated system in terms of the hierarchy
of reduced distribution functions. Section {\bf III}, is devoted to
developing the thermodynamic analysis, and to deriving the entropy
production. Here, we draw kinetic equations as one of the consequences, in
particular the Boltzmann equation. In section {\bf IV}, we stress our main
conclusions.

{\em Distribution vector dynamics}.---Let us think of a dynamical system
having N interacting degrees of freedom $%
(q_{1},......,q_{2},p_{1},......,p_{N})$, where $q_{i}$ and $p_{i}$ are the
generalized coordinate and conjugated momentum corresponding to the i-th
degree of freedom. Let $H(\left\{ q^{N},p^{N}\right\} )$ be the Hamiltonian
of the system, given by 
\begin{equation}
H=\sum_{j=1}^{N}\left\{ \frac{p_{j}^{2}}{2m}+\phi ^{e}(q_{j})\right\} +\frac{%
1}{2}\sum_{j\neq k=1}^{N}\phi \left( \left| q_{j}-q_{k}\right| \right) \text{
,}  \label{hamiltonian}
\end{equation}
where $m$ is the mass, $\phi ^{e}(q_{j})\equiv \phi ^{e}{}_{j}$ is an
external potential, and $\phi \left( \left| q_{j}-q_{k}\right| \right)
\equiv \phi _{jk}$ is the interaction potential. The state of the system is
completely specified at a given time by the N-particle distribution function 
$F\left( \left\{ q^{N},p^{N}\right\} ;t\right) $, which evolves in time
according to the Liouville equation. Nonetheless, an alternative description
of the state of the system can be given in terms of the distribution vector 
\cite{balescu} 
\begin{equation}
{\bf f}\equiv \left\{
f_{o},f_{1}(x_{1}),f_{2}(x_{1},x_{2}),.........,f_{N}(x_{1},x_{2,}......,x_{N})\right\} 
\text{ ,}  \label{distributionvector}
\end{equation}
which represents the set of all the reduced s-particle distribution
functions $f_{s}(x_{1},x_{2,}......,x_{s})$ $(s=0,........,N)$, defined
through 
\begin{equation}
f_{s}=\frac{N!}{(N-s)!}\int F(x_{1},.....,x_{s},x_{s+1},......,x_{N})\text{ }%
dx_{s+1}.....dx_{N}\text{ ,}  \label{reduceddistribution}
\end{equation}
where $x_{j}\equiv (q_{j},p_{j})$. This function is normalized according to 
\begin{equation}
\int f_{s}\text{ }dx_{1}.....dx_{s}=\frac{N!}{(N-s)!}\text{ \ .}
\label{normalization}
\end{equation}
The evolution equations of the s-particle distribution functions can be
obtained by integrating the Liouville equation constituting a set of coupled
equations: the BBGKY hierarchy.

An arbitrary component equation of this hierarchy is 
\begin{eqnarray}
\frac{\partial }{\partial t}f_{s}(x_{1},......,x_{s}) &=&\left(
\sum_{j=1}^{s}L_{j}^{o}+\sum_{j<n=1}^{s}L_{j,n}^{\prime }\right)
f_{s}(x_{1},....,x_{s})  \nonumber \\
&&+\sum_{j=1}^{s}\int L_{j,s+1}^{\prime }f_{s+1}(x_{1},....,x_{s+1})\text{ }%
dx_{s+1}\text{,}  \label{hierarchy}
\end{eqnarray}
which shows that the time derivative of $f_{s}$ depends both on $f_{s}$ and
on the higher-order functions $f_{s+1}$. Here 
\begin{equation}
L_{j}^{o}=\left[ H_{j}^{o},...\right] _{P}\text{ ,}  \label{poisson0}
\end{equation}
where $\left[ ...,...\right] _{P}$ is the Poisson bracket, $H_{j}^{o}=\frac{%
p_{j}^{2}}{2m}+\phi _{j}^{e}$, and 
\begin{equation}
L_{j,n}^{\prime }=\left[ H_{j,n}^{\prime },...\right] _{P}\text{ ,}
\label{poisson1}
\end{equation}
with $H_{j,n}^{\prime }=\frac{1}{2}\phi _{j,n}$.\qquad

From Eq. (\ref{hierarchy}) one can infer the evolution equation for the
distribution vector 
\begin{equation}
\frac{\partial }{\partial t}{\bf f}(t)={\cal L}{\bf f}(t)\text{ ,}
\label{generalizedliouville}
\end{equation}
which constitutes the generalized Liouville equation, and ${\cal L}$ is the
generalized Liouvillian. This equation should be equivalent to Eq. (\ref
{hierarchy}), thus, the matrix elements $\langle s\left| {\cal L}\right|
s^{\prime }\rangle $ must have the following values: 
\begin{eqnarray}
\langle s\left| {\cal L}\right| s^{\prime }\rangle  &=&\delta _{s,s^{\prime
}}\left\{ \sum_{j=1}^{s}L_{j}^{o}+\sum_{j<n=1}^{s}L_{j,n}^{\prime }\right\} 
\nonumber \\
&&+\delta _{s^{\prime },s+1}\int \left\{ \sum_{j=1}^{s}L_{j,s+1}^{\prime
}\right\} dx_{s+1}\text{ ,}  \label{matrixelement}
\end{eqnarray}
and 
\begin{equation}
\langle 0\left| {\cal L}\right| s^{\prime }\rangle =0\text{ ,}
\label{matrixelementnul}
\end{equation}
where $\mid s\rangle $ represents the s-particle state. At this point, it is
convenient to introduce the operator ${\cal P}$ providing the diagonal part
of a matrix , and ${\cal Q}$, its complement which gives the nondiagonal
part. Thus, by using these operators we can rewrite Eq. (\ref
{generalizedliouville}) as follows 
\begin{equation}
\left( \frac{\partial }{\partial t}-{\cal PL}\right) {\bf f}(t)={\cal QL}%
{\bf f}(t)\text{ ,}  \label{separatedgeneralizedliouville}
\end{equation}
with 
\begin{equation}
\langle s\left| {\cal PL}\right| s^{\prime }\rangle =\delta _{s,s^{\prime
}}\left\{ \sum_{j=1}^{s}L_{j}^{o}+\sum_{j<n=1}^{s}L_{j,n}^{\prime }\right\} 
\text{ ,}  \label{diagonal}
\end{equation}
and 
\begin{equation}
\langle s\left| {\cal QL}\right| s^{\prime }\rangle =\delta _{s^{\prime
},s+1}\int \left\{ \sum_{j=1}^{s}L_{j,s+1}^{\prime }\right\} dx_{s+1}\text{ .%
}  \label{nondiagonal}
\end{equation}
The left-hand side of Eq. (\ref{separatedgeneralizedliouville}) is invariant
under time reversal whereas this symmetry is broken in the right-hand side
term, as one can convince oneself changing $t$ by $-t$ and taking into
account Eqs. (\ref{diagonal}) and (\ref{nondiagonal}). Thus, the diagonal
part of the generalized Liouvillian accounts for the reversible evolution of
the distribution vector, and the nondiagonal part introduces irreversibility
in the dynamics of ${\bf f}$ leading to dissipation, this fact will be show
adequately in the section that follows. Hence, in the sense that the time
reversal invariance is broken, the generalized Liouville equation given
through Eqns. (\ref{generalizedliouville}) or (\ref
{separatedgeneralizedliouville}) is irreversible. Irreversibility is
manifested in the dynamics of the system when we use the adequate
description, {\it i.e. }in terms of the distribution vector, this
description brings with itself certain degree of coarse graining.{\em \ }

{\em Mesoscopic nonequilibrium thermodynamics}.--- The MNET constitutes the
mesoscopic generalization of the macroscopic formalism of the
non-equilibrium thermodynamics \cite{degroot}. Here, the entropy is a
functional of the distribution vector in the phase space given by the Gibbs
entropy postulate \cite{vankampen}. Moreover, we apply the principles of
equilibrium thermodynamics at a local level in phase space, and we follow
the general scheme of non-equilibrium thermodynamics in order to compute the
entropy production from the Gibbs entropy postulate. One then postulates the
linear phenomenological equations relating fluxes and thermodynamic forces
occurring therein. Immediately afterwards, by substituting this
phenomenological equations in the generalized Liouville equation written in
the form given in Eq. (\ref{separatedgeneralizedliouville}), one derives
master equations for the distribution vector. Therefore, the entropy of the
system, in analogy to the Gibbs entropy postulate, is given by 
\begin{eqnarray}
S &=&-k_{B}Tr\left\{ {\bf f}\log \left( {\bf f}_{o}^{-1}{\bf f}\right)
\right\} +S_{o}  \nonumber \\
&=&-k_{B}\sum_{s=1}^{N}\int f_{s}\log \frac{f_{s}}{f_{os}}\;dx_{1}.....dx_{s}%
\text{ }+S_{o}\text{ ,}  \label{gibbspostulate}
\end{eqnarray}

where $k_{B}$ is the Boltzmann constant, $S_{o}$ the equilibrium entropy,
and ${\bf f}_{o}$ is assumed to be the equilibrium distribution vector
satisfying ${\cal L}{\bf f}_{o}=0$. The rate of change of $S$ can be
obtained by differentiating Eq. (\ref{gibbspostulate}) 
\begin{equation}
\frac{dS}{dt}=-k_{B}Tr\left\{ \frac{\partial {\bf f}}{\partial t}\log \left( 
{\bf f}_{o}^{-1}{\bf f}\right) \right\} \text{ ,}  \label{entropyproduction}
\end{equation}
which for small deviations from the equilibrium reduces to
\begin{equation}
\frac{dS}{dt}=-k_{B}Tr\left\{ \frac{\partial {\bf f}}{\partial t}{\bf f}%
_{o}^{-1}\left( {\bf f}_{o}-{\bf f}\right) \right\} \text{ .}
\label{linearentropyproduction}
\end{equation}
Here, the only contribution to the entropy production comes from the
irreversible change of the distribution vector since the change reversible
contributes to zero entropy production. The reversible change of the
distribution vector is given by the mainstream current induced by the
s-Hamiltonian flow in each one of the s-particle subspaces of the total
phase space, and it is known that the Hamiltonian flow does not create
entropy. According to the Onsager theory \cite{degroot}, \cite{mazurbedeaux}
from Eq. (\ref{linearentropyproduction}) we infer the linear relation 
\begin{equation}
\frac{\partial {\bf f}}{\partial t}={\bf M}\left\{ {\bf f}_{o}^{-1}\left( 
{\bf f}_{o}-{\bf f}\right) \right\} \equiv {\bf MX}\text{ ,}
\label{linearelation}
\end{equation}
where the master or mobility matrix ${\bf M}$ acts on an arbitrary vector $%
{\bf Y}$ according to 
\begin{equation}
\langle s\left| {\bf M}\right| s^{\prime }\rangle \langle s^{\prime }\mid 
{\bf Y=}\int M{\bf (}s{\bf \mid }s^{\prime })Y_{s^{\prime }}(x_{1}^{\prime
},....,x_{s^{\prime }}^{\prime })dx_{1}^{\prime }.....dx_{s^{\prime
}}^{\prime }\text{ .}  \label{mobilitycomponents}
\end{equation}
Substituting the linear relation Eq. (\ref{linearelation}) into Eq. (\ref
{entropyproduction}) and using the cyclic invariance of the trace, we find
for the entropy production 
\begin{equation}
Tr\left( {\bf XMX}\right) \geq 0\text{ ,}  \label{secondlaw}
\end{equation}
according to the second law of Thermodynamics. This relation imposes the
Hermitian character of the master matrix 
\begin{equation}
M{\bf (}s{\bf \mid }s^{\prime })=M{\bf (}s^{\prime }{\bf \mid }s)^{\dagger }%
\text{ ,}  \label{onsager}
\end{equation}
as predicted by the Onsager symmetry relations. Here $^{\dagger }$ refers to
the Hermitian conjugate. Furthermore, due to the fact that ${\bf f}$ should
be normalized $Tr({\bf f)}$ is a constant quantity which is a function of $N$%
, from Eq. (\ref{linearelation}) we infer that $Tr({\bf MX)}=0$, which, when
we take into account the Hermitian character of the master matrix, leads to
the following constraints 
\begin{equation}
\int M(s\mid s^{\prime })dx_{1}.....dx_{s}=\int M(s\mid s^{\prime
})^{^{\dagger }}dx_{1}^{\prime }.....dx_{s^{\prime }}^{\prime }=0\text{ .}
\label{constrains}
\end{equation}

Hence, by using the generalized Liouville equation (\ref
{generalizedliouville}) and Eq. (\ref{linearelation}) it comes to be that
Eq. (\ref{separatedgeneralizedliouville}) can be written in a more
convenient form 
\begin{equation}
\left( \frac{\partial }{\partial t}-{\cal PL}\right) {\bf f}(t)={\cal Q}%
\text{ }\frac{\partial {\bf f}}{\partial t}={\cal Q}{\bf MX}\text{ ,}
\label{generalizedkinetic}
\end{equation}
with 
\begin{eqnarray}
\langle s\left| {\cal Q}{\bf M}\right| s^{\prime }\rangle \langle s^{\prime
} &\mid &{\bf X}  \nonumber \\
&=&\delta _{s^{\prime },s+1}\int M{\bf (}s+1{\bf \mid }s^{\prime
})X_{s^{\prime }}dx_{1}^{\prime }.....dx_{s^{\prime }}^{\prime }dx_{s+1}%
\text{ ,}  \label{projectionm}
\end{eqnarray}
written in analogy with Eq. (\ref{nondiagonal}). It should be mentioned that
the presence of the master matrix in Eq. (\ref{generalizedkinetic}) notably
simplifies the BBGKY hierarchy. In fact, the master matrix introduces a
relaxation time scale. Hence, our theory is equivalent to a relaxation time
approach to the study of the BBGKY hierarchy valid when there is a broad
separation between the hydrodynamic and microscopic scales.

Here, Eq. (\ref{generalizedkinetic}) can be brought into a more common form,
by introducing a new function ${\bf W}$ defined through \cite{meixner} 
\begin{eqnarray}
f_{o,s^{\prime }}W\left( s^{\prime }\mid s\right) &=&-M\left( s\mid
s^{\prime }\right)  \nonumber \\
&&+\delta \left( x_{1}^{\prime }-x_{1}\right) ......\delta \left(
x_{s^{\prime }}^{\prime }-x_{s}\right) \psi _{s}\left(
x_{1},....,x_{s}\right) \text{ ,}  \label{transitionmatrix}
\end{eqnarray}
where the auxiliary function $\psi _{s}$ is not arbitrarily selectable
because of the constraints given by Eq. (\ref{constrains}). Rather applies 
\begin{equation}
\psi _{s}\left( x_{1},.....,x_{s}\right) =f_{o,s}(x_{1},......,x_{s})\int W%
{\bf (}s{\bf \mid }s^{\prime })dx_{1}^{\prime }.....dx_{s^{\prime }}^{\prime
}\text{ .}  \label{auxiliaryfunction}
\end{equation}
Thus, Eq. (\ref{generalizedkinetic}) may be written in terms of the
transition matrix as follows 
\begin{equation}
\left( \frac{\partial }{\partial t}-{\cal PL}\right) {\bf f}(t)={\cal Q}{\bf %
Wf}(t)\text{ ,}  \label{geralizedkinetic2}
\end{equation}
where

\begin{eqnarray}
\langle s\left| {\cal Q}{\bf W}\right| s^{\prime }\rangle \langle s^{\prime
} &\mid &{\bf f=}\delta _{s^{\prime },s+1}\int \{f_{s^{\prime
}}(x_{1}^{\prime },.......,x_{s^{\prime }}^{\prime })W{\bf (}s^{\prime }{\bf %
\mid }s+1)-  \nonumber \\
f_{s+1}(x_{1},.......,x_{s+1})W{\bf (}s+1 &\mid &s^{\prime
})\}dx_{1}^{\prime }.....dx_{s^{\prime }}^{\prime }dx_{s+1}\text{ .}
\label{dissipationterm}
\end{eqnarray}
It should be stressed that the introduction of the transition matrix leads
to a deeper physical insight of our result.

A particularly interesting case corresponds to $s=1$ where Eqs. (\ref
{geralizedkinetic2}) and (\ref{dissipationterm}) reduce to 
\begin{eqnarray}
\left( \frac{\partial }{\partial t}-L_{1}^{o}\right) f_{1}(x_{1}) &=&\int
\{f_{2}(x_{1}^{\prime },x_{2}^{\prime })W{\bf (}x_{1}^{\prime
},x_{2}^{\prime }{\bf \mid }x_{1},x_{2})  \nonumber \\
-f_{2}(x_{1},x_{2})W{\bf (}x_{1},x_{2} &\mid &x_{1}^{\prime },x_{2}^{\prime
})\}dx_{1}^{\prime }dx_{2}^{\prime }dx_{2}\text{ .}  \label{preboltzmann}
\end{eqnarray}
This is not yet a kinetic equation; however, its uncorrelated part is and
can be written as 
\begin{eqnarray}
\left( \frac{\partial }{\partial t}-L_{1}^{o}\right) f_{1}(x_{1}) &=&\int
\{f_{1}(x_{1}^{\prime })f_{1}(x_{2}^{\prime })W{\bf (}x_{1}^{\prime
},x_{2}^{\prime }{\bf \mid }x_{1},x_{2})  \nonumber \\
-f_{1}(x_{1})f_{1}(x_{2})W{\bf (}x_{1},x_{2} &\mid &x_{1}^{\prime
},x_{2}^{\prime })\}dx_{1}^{\prime }dx_{2}^{\prime }dx_{2}\text{ ,}
\label{boltzmann}
\end{eqnarray}
which is the famous Boltzmann equation. The correlated part of Eq. (\ref
{preboltzmann}) does not matter because the correlated part of each side of
this equation is identically zero. Analogous procedure can be followed to
obtain the kinetic equations for the correlations coming from the components
of ${\bf f}$ of an order higher than one.

The quantum generalization of our classical treatment can be done in terms
of the Wigner distribution vector \cite{balescu}, defined as the set of all
the s-particle Wigner functions. This quantum approach follows the same
lines as the previous classical approach, but with the unavoidable
differences in the algebraic manipulations.

{\em Discussion}.---Here, we have presented a representation of the
statistical description of a many-body system in terms of the distribution
vector, which reveals the irreversible component of the motion of this
system. This component is hidden when we work with the standard Liouville
equation but at the level of the generalized Liouville equation which
entails a certain degree of coarse graining, this irreversible component
becomes obvious. The Liouville equation is a closed equation for the phase
space distribution function while the generalized Liouville equation
encloses a set of coupled equations, each one in itself represents a
contraction of the statistical description. But, as we have mentioned in
section II, the full statistical description in terms of the phase space
distribution function is equivalent to the description in terms of the set
of all the reduced s-particle distribution functions. In this scenario, we
show that the Gibbs entropy given by the Gibbs entropy postulate is a
non-conserved quantity throughout the motion in phase space. This entropy
should increase until its maximum value at equilibrium according to the
second law. The consequences we have drawn from the Gibbs entropy are
compatible with the explanation of irreversibility which follows from
Boltzmann's entropy postulate.

By applying Onsager's theory in phase space, {\it i.e.} in the framework of
MNET, we obtain a linear relation between the irreversible rate of change of
the distribution vector and the conjugated thermodynamic force, which
introduces a master matrix which should satisfy the properties required by
the Onsager theory. This result simplifies the BBGKY hierarchy and enables
us to break the hierarchy obtaining the kinetic equation for the
one-particle distribution function which coincides with the Boltzmann
kinetic equation. Hence, this last result constitutes a test of our theory.
Hence, we have shown a method based on the MNET for deriving kinetic
equations.

We also stress that in our mesoscopic approach, the master matrix introduces
a relaxation time scale which shows that our theory is valid when there is a
broad separation between the hydrodynamic and microscopic scales. This
separation of time scales allows the rate of change of the distribution
vector due to the main stream flow velocity in phase space balance the
irreversible change in order to achieve a stationary distribution at long
time. This is precisely what happens in the context of the phenomena
described by the kinetic equations.

I want acknowledge Prof. J.M. Rub\'{\i} for profitable discussions. This
work was supported by DGICYT of the Spanish Governement under Grant No.
PB2002-01267.

\end{document}